\documentclass[universe,article,submit,pdftex,moreauthors]{mdpi} 

\firstpage{1} 
\pubvolume{1}
\issuenum{1}
\articlenumber{0}
\pubyear{2022}
\copyrightyear{2022}
\datereceived{} 
\dateaccepted{} 
\datepublished{} 
\hreflink{https://doi.org/} 

\usepackage{graphicx}

\usepackage{lscape}
\usepackage{longtable}
\usepackage{verbatim}
\Title{The Classification of Blazars Candidates of Uncertain Types}
\TitleCitation{Classification of BCUs}

\Author{Jun-Hui Fan $^{1,2,3}$*, Ke-Yin Chen $^{4}$, Hu-Bing Xiao $^{5}$**, Wen-Xin Yang $^{1,2,3}$,
Jing-Chao Liang $^{1,2,3}$, Guo-Hai Chen $^{1,2,3}$, Jiang-He Yang $^{6,1}$, Yu-Hai Yuan $^{1,2,3}$, De-Xiang Wu $^{1,2,3}$}

\AuthorNames{Jun-Hui Fan, Ke-Yin Chen, Hu-Bing Xiao, Wen-Xin Yang, Jing-Chao Liang, Guo-Hai Chen, Jiang-He Yang, Yu-Hai Yuan and De-Xiang Wu}

\AuthorCitation{Fan, J.-H.; Chen, K.-Y.; Xiao, H.-B.; Yang, W.-X.; Liang, J.-C.; Chen. G.-H,; Yang, J.-H.; Yuan, Y.-H.; Wu, D.-X.}

\address{%
$^{1}$ \quad Center for Astrophysics, Guangzhou University, WaiHuanXi Road No. 230, Guangzhou 510006, China; fjh@gzhu.edu.cn\\
$^{2}$ \quad Astronomy Science and Technology Research Laboratory of Department of Education of Guangdong Province, WaiHuanXi Road No. 230, Guangzhou 510006, China\\
$^{3}$ \quad Key Laboratory for Astronomical Observation and Technology of Guangzhou, WaiHuanXi Road No. 230, Guangzhou 510006, China\\
$^{4}$ \quad Department of Physics, School of Physics and Material Science, Guangzhou University, WaiHuanXi Road No. 230, Guangzhou 510006, China\\
$^{5}$ \quad Shanghai Key Lab for Astrophysics, Shanghai Normal University, GuiLin Road No.100, Shanghai 200234, China\\
$^{6}$ \quad Department of Physics and Electronics Science, Hunan University of Arts and Science, DongTing Avenue No. 3150, Changde 415000, China\\}

\corres{Correspondence: fjh@gzhu.edu.cn; hubing.xiao@shnu.edu.cn}
\preto{\abstractkeywords}{\nolinenumbers}

\abstract{
 In this work,
  the support vector machine (SVM) method is adopted to
  separate BL Lacertae objects (BL Lacs) and flat spectrum radio quasars (FSRQs)
  in the plots of
  photon  spectrum index against the photon flux,
  $\alpha_{\rm ph} \sim {\rm log}\,F$,
  that of
  photon spectrum index against the variability index,
  $\alpha_{\rm ph} \sim {\rm log}\, \textit{V\!I}$, and
  that of
  variability index against the photon flux,
  ${\rm log}\,{V\!I} \sim {\rm log}\,F$.
  Then we used the dividing lines to tell BL Lacs from FSRQs in
  the blazars candidates of uncertain types from \textit{Fermi}/LAT catalog.
  Our main conclusions are:
  1. We separate BL Lacs and FSRQs by
     $\alpha_{\rm ph} = -0.123\,{\rm log}\,F + 1.170$
     in the $\alpha_{\rm ph} \sim {\rm log}\,F$ plot,
     $\alpha_{\rm ph} = -0.161\,{\rm log}\,{V\!I} + 2.594$
     in the $\alpha_{\rm ph} \sim {\rm log}\,{V\!I}$ plot, and
      ${\rm log}\,{V\!I} = 0.792\,{\rm log}\,F + 9.203$
     in the ${\rm log}\,{V\!I} \sim {\rm log}\,F$ plot.
   2. We obtained
        932 BL Lac candidates and possible BL Lac candidates,
        and 585 FSRQ candidates and possible FSRQ candidates.
   3. Some discussions are given for the comparisons with those in the literatures.}

\begin{document}

\section{Introduction}           
\label{sect:intro}

  As a special subclass of  active galactic nuclei (AGNs),
  blazars show some extreme observational properties:
    high amplitude and rapid variability superposed on the
    long-term slow variation light curve,
    high polarization,
    powerful $\gamma$-ray emissions,
    some sources emitting TeV emissions,
    strong broad emission line features or have no emission line at all,
    or superluminal motions,
    \citep{
    Abd20,
    acero15,
    aje20,
fan21,
ghi14,
    sti91,
    urry95,
    Wills92,
Yang2022b,
zhou21,
Villata2006,
Gupta2016,
Lister2018,
Lister2021}.
    Those extreme observational properties are explained by
    a beaming model,
    in which  there is a central supermassive black hole surrounded by an accretion disk,
    and two jets being perpendicular to the disk.
    When the jet points to the observer,
    the observed emission, $f^{\rm ob}$, is boosted and
    the variability time scale, $\Delta t^{\rm ob}$ is shortened by
        $f^{\rm ob}=\delta^q f^{\rm in}$, and
        $\Delta t^{\rm ob} = \Delta t^{\rm in}/ \delta$,
    where
    $f^{\rm in}$ and $\Delta t^{\rm in}$ are the emission and
    the variability time scale in the comoving frame,
    $\delta=[\Gamma(1-\beta \cos \theta)]^{-1}$ is a boosting factor (or Doppler factor),
    $\Gamma=1 / \sqrt{1-\beta^{2}}$, is the Lorentz factor,
     $\theta$ is the viewing angle, and
    $\beta$ is the jet speed in units of the speed of light,
    $\beta = v/c$, and $q$ is a parameter depending on the jet case:
        $q = 2 + \alpha$ is for  the case of a continuous jet,
        while $q = 3 + \alpha$ is for the case of a moving sphere
        \citep{Lind85}, and
    $\alpha$ is a spectral index defined by $f_{\nu} \propto \nu^{-\alpha}$.

  Based on the behavior of emission lines,
  blazars can be classified into two subclasses,
  namely BL Lacertae objects (BL Lacs) and
   flat spectrum radio quasars (FSRQs).
  FSRQs have strong broad emission lines with the equivalent width being greater than 5 $\AA$ ($\textit{E\!W} > 5 \AA$),
   while BL Lacs show only weak or no emission line at all, or $\textit{E\!W} < 5 \AA$ \citep{Stocke1990, sti91, Scarpa1997}.
   BL Lacs were classified as radio  selected BL Lacs (RBLs)
   and X-ray selected BL Lacs (XBLs) from surveys.
   The both have differently observational properties in
   Hubble diagrams,
  multiwavelength correlations,
  spectral  index correlations,
  and linear optical polarizations, etc. \citep{Fan96} and references therein.
   The physical classification of BL Lacs was that by
   \citet{pg95}, who calculated the spectral energy distributions (SEDs) for
   a sample of BL Lacs objects and proposed to use the synchrotron peak
   frequency ($\nu_{\rm p}$) to separate BL Lacs into
   highly peaked BL Lacs (HBLs) with ${\rm log}\,\nu_{\rm p}\,{\rm (Hz)} > 15$ (the base of the logarithms is ten throughout this paper) and
   lowly peaked BL Lacs (LBLs) with ${\rm log}\,\nu_{\rm p}\,{\rm (Hz)} < 15$, and
   it was found that most RBLs belong to LBLs while
   XBLs to HBLs.

   \citet{nie06}  calculated SEDs, ${\rm log}\,\nu F_{\nu} - {\rm
log} \,\nu $ for 308 BL Lacs, classified them into three subclasses:
LBLs, IBLs, and HBLs with the boundaries being ${\rm log}\,\nu_{\rm p}\, {\rm (Hz)} = 14.5$ and ${\rm log} \,\nu_{\rm p}\, {\rm (Hz)} = 16.5$ respectively.
So, it was set as
 LBLs if  ${\rm log}\,\nu_{\rm p}\, {\rm (Hz)} < 14.5$,
 IBLs if $14.5 < {\rm log}\,\nu_{\rm p}\, {\rm (Hz)} < 16.5$, and
 HBLs if ${\rm log}\,\nu_{\rm p}\, {\rm (Hz)} > 16.5$ for HBLs.
 Later on,
 \citet{abdo10}
  calculated SEDs using the quasi-simultaneous data of 48 Fermi blazars,
  and extended the definition to all types of non-thermal dominated AGNs
  using new acronyms:
  low synchrotron peaked blazars(LSP)
    if  ${\rm log}\,\nu_{\rm p}\, {\rm (Hz)} < 14$,
  intermediate synchrotron peaked blazars (ISP)
    if $14  < {\rm log}\,\nu_{\rm p} \, {\rm (Hz)}  < 15$, and
  high synchrotron peaked blazars (HSP),
    if ${\rm log}\,\nu_{\rm p}\, {\rm (Hz)}  > 15$.
  They also proposed an empirical parametrization to estimate the synchrotron peak frequency
  using the effective  radio-optics spectral index ($\alpha_{\rm ro}$) and
   the effective optics-X-ray spectral index ($\alpha_{\rm ox}$).

   Following
   \citet{abdo10} and using their acronyms,
   \citet{fan16} calculated the SEDs for 1392 \textit{Fermi} blazars
  and proposed  following classifications:
   LSPs if ${\rm log}\, \nu_{\rm p}\, ({\rm Hz}) \leq 14.0$,
  ISPs if $14.0 < {\rm log}\, \nu_{\rm p}\, ({\rm Hz}) \leq 15.3$, and
  HSPs if ${\rm log}\, \nu_{\rm p}\, ({\rm Hz}) > 15.3$.
  They found no component with extreme high peak frequency of ${\rm log}\, \nu_{\rm p}\, ({\rm Hz}) > 19$.

  Very recently,
  \citet{Yang2022b} calculated  the SEDs for 2709 blazars (including BCUs) in 4FGL DR3
  and obtained following classifications:
   LSPs if ${\rm log}\, \nu_{\rm p}\, ({\rm Hz}) \leq 13.7$,
  ISPs if $13.7 < {\rm log}\, \nu_{\rm p}\, ({\rm Hz}) \leq 14.9$, and
  HSPs if ${\rm log}\, \nu_{\rm p}\, ({\rm Hz}) > 14.9$.
  These blazar classification boundaries are summarized in Table \ref{bou}.

\begin{table*}
\scriptsize \caption{The boundary of blazar classification}
\label{bou} \centering
\begin{tabular}{lccccc}
\hline \hline
Type & Lower & Intermediate & Higher & Ref. & $N$ \\ \hline
BL Lacs 	& ${\rm log}\, \nu_{\rm p}\, ({\rm Hz}) < 14.5$ 	& 	$14.5  < {\rm log}\,\nu_{\rm p} \, {\rm (Hz)}  < 16.5$ 	&	${\rm log}\,\nu_{\rm p} \, {\rm (Hz)}  > 16.5$	&	\citet{nie06}		&	308 		\\ \hline
		& ${\rm log}\, \nu_{\rm p}\, ({\rm Hz}) < 14.0$ 	& 	$14.0  < {\rm log}\,\nu_{\rm p} \, {\rm (Hz)}  < 15.0$ 	&	${\rm log}\,\nu_{\rm p} \, {\rm (Hz)}  > 15.0$	&	\citet{abdo10}		&	48 		\\ 
Blazars	& ${\rm log}\, \nu_{\rm p}\, ({\rm Hz}) < 14.0$ 	& 	$14.0  < {\rm log}\,\nu_{\rm p} \, {\rm (Hz)}  < 15.3$ 	&	${\rm log}\,\nu_{\rm p} \, {\rm (Hz)}  > 15.3$	&	\citet{fan16}		&	1392 	\\ 
		& ${\rm log}\, \nu_{\rm p}\, ({\rm Hz}) < 13.7$ 	& 	$13.7  < {\rm log}\,\nu_{\rm p} \, {\rm (Hz)}  < 14.9$ 	&	${\rm log}\,\nu_{\rm p} \, {\rm (Hz)}  > 14.9$	&	\citet{Yang2022b}	&	2709 	\\
\hline
\end{tabular}
\\ *$N$ denotes the sample size that the authors used to get the boundary.
\end{table*}

   \textit{Fermi}/LAT missions have detected a lot of $\gamma$-ray emitters,
   more than 60\% of the \textit{Fermi}/LAT detected sources are AGNs.
   and 99\% of the \textit{Fermi}/LAT AGNs are blazars
   So, $\gamma$-ray emission is a typically observational property of blazars,
   and $\gamma$-ray emission was taken as one of the observation
   properties of blazars.
   Up to now,
   several catalogues have been released
   \citep{
   abdo10,
   Nolan12,
   acero15,
   aje20,
   Ballet2020}.

  The $\gamma$-ray loud blazars are variable on different time scales \citep{Yang2022b}.
  \textit{Fermi}/LAT detected a lot of $\gamma$-ray emitters,
  there are 5 catalogues of \textit{Fermi}/LAT mission,
  which provide us with a nice opportunity to investigate the
  variability properties in the $\gamma$-ray band.
  The variability level in the $\gamma$-ray was introduced by a
  so called variability index (${V\!I}$)  defined as by \citet{Abd20}:
\begin{equation}
{V\!I} = 2 \sum_{i} {\rm log}\, \left[ \frac{\mathcal{L}_{i}(S_{i})}{\mathcal{L}_{i}(S_{\rm glob})} \right] - {\rm max({\chi}^{2}(S_{\rm glob}) - {\chi}^{2}(S_{\rm av}), 0)},
\end{equation}
\begin{equation}
\chi^{2}(S) = \sum_{i} \frac{(S_{i}-S)^{2}}{\sigma_{i}^{2}},
\end{equation}
 here $S_{i}$ are the individual flux values, $\mathcal{L}_{i}(S)$ is the likelihood in the interval $i$ assuming flux $S$, and $\sigma_{i}$ are the errors on $S_{i}$, $S_{\rm av}$ is the average flux and $S_{\rm glob}$ is the globe flux.

  The latest fourth \textit{Fermi}/LAT catalog (4FGL) with 5099
  sources was published \citep{Abd20,aje20}.
  Out of them
  1432 are BL Lacs,
  795 are FSRQs, and
  1518 blazar candidates of uncertain type (BCUs).
  The identification of the BCUs is interesting and it can provide more sources
  for us to investigate the different physics in BL Lacs and FSRQs.
  The identifications of BCUs were carried out in many works
\citep{Hassan2013, Doert2014, Chiaro2016, Parkinson2016, Lefaucheur2017, Yi2017, Bai2018, Ma2019, Kang2019, kang19, Yang2022b}

  In this work,
  we apply the support vector machine (SVM) learning method to separate BL Lacs and FSRQs
  and then use the dividing line to tell BL Lac candidates from FSRQ candidates from the BCUs.
  The work is arranged as follows:
  In the 2nd section,
  a sample, from 4FGL\_DR3,  used in the work will be described,
  in 3rd section the distributions of the physical parameters will be
  given for BL Lacs and FSRQs, and the SVM method will be used to separate BL Lacs and FSRQs,
  and divide BL Lac candidates and FSRQ candidates,
  some discussions and conclusions are given in section 4 and section 5.

\section{Sample and  Classifications}
\subsection{Samples}

In this work,
we obtained  3743 blazars from  the 4FGL catalogue
 \citep{Abd20,aje20},  which include
  1432 BL Lacs,
  794 FSRQs, and
  1517 BCUs.
  We only list 1517 BCUs in Table \ref{Com} since we want to
  classify them in this work.

  Redshift, which is from NEDs (https://ned.ipac.caltech.edu/classic/),
  is available for 2094 sources (1010 BL Lacs, 781 FSRQs and 303 BCUs).
  The redshift is in a range  of
  $z  = $ 0.0005 for 4FGL J1124.0+2045
   to $z = $ 5.540 for 4FGL J1113.9+5523  for the 2094 sources.
  The averaged values are
  $\langle z \rangle = 0.507 \pm 0.512$ for 1010 BL Lacs,
  $\langle z \rangle = 1.200 \pm 0.662$ for 781 FSRQs, and
  $\langle z \rangle = 0.777 \pm 0.710$ for 303 BCUs.

\subsection{Average Values}

\begin{figure}
\begin{center}
\includegraphics[angle=0,width=13cm]{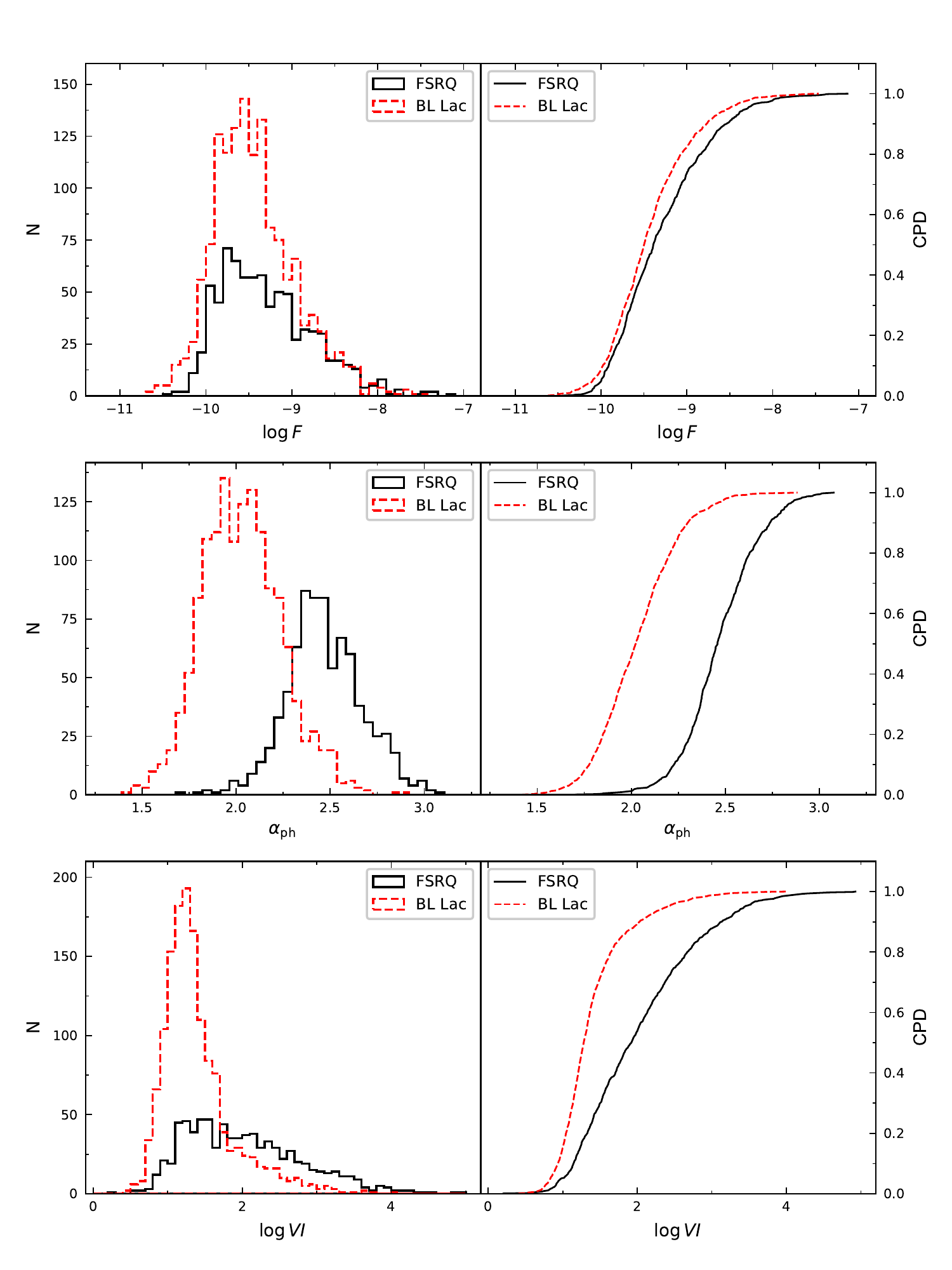}
\caption{Histograms (left panel) for FSRQs and BL Lacs and
their corresponding cumulative probability distribution (CPD, right panel) for three parameters.
In this plot, the dashed red line for the BL Lacs, and the solid black line for the FSRQs.
Upper panel:  for logarithm of the $\gamma$-ray photons, ${\rm log}\,F$ in units of ph/cm$^2$/s,
 middle panel:  for the $\gamma$-ray photon spectral index, $\alpha_{\rm ph}$,
 bottom panel:  for logarithm of the $\gamma$-ray variability index, ${\rm log}\,{V\!I}$,
 } \label{Hist-KS}
\end{center}
\end{figure}

\textbf{\textit{$\gamma$-Ray Photon Flux - ${\rm log}\,F$}}:
Based on the photon flux intensity from the 4FGL catalogue \citep{Abd20,aje20},
we got the logarithm of the $\gamma$-ray photon flux (${\rm log}\,F$) and showed their distributions for FSRQs and BL Lacs in the upper-left panel of
Fig. \ref{Hist-KS},
and their cumulative distributions are in the upper-right panel of Fig. \ref{Hist-KS}.
Their averaged values are
$\langle {\rm log}\,F \rangle \, = -9.294\, \pm\, 0.520$ for FSRQs, and
$\langle {\rm log}\,F \rangle\, = -9.434\, \pm\, 0.482$  for BL Lacs.
When a  K-S test is performed to the distributions,
a probability $p = 6.708\times10^{-7}$ for the two distributions
to be  from the same parent  distribution is obtained.

\textbf{\textit{Photon Spectral  Index-$\alpha_{\rm ph}$}}:
We showed the distributions of $\alpha_{\rm ph}$ for FSRQs and BL Lacs in the middle-left panel in Fig. \ref{Hist-KS}, and their cumulative distributions are showed in the middle-right panel of Fig. \ref{Hist-KS}.
The average photon spectral indexes are
$\langle \alpha_{\rm ph} \rangle \, = 2.470\, \pm\, 0.201 $ for 795 FSRQs, and
$\langle \alpha_{\rm ph} \rangle \, = 2.032\, \pm\, 0.212 $ for 1432 BL Lacs.
The K-S test gives  $p = 7.77\times10^{-16}$.

\textbf{\textit{Variability Index - {V\!I}}}: 
For the variability index, we calculated the corresponding logarithm and showed their distributions for FSRQs and BL Lacs in the lower-left panel of Fig. \ref{Hist-KS},
and their cumulative distributions are in the lower-right panel of Fig. \ref{Hist-KS}.
For the averaged values, we have
$\langle {\rm log}\, {V\!I} \rangle \, = 2.025\, \pm\, 0.777$ for FSRQs, and
$\langle {\rm log}\, {V\!I} \rangle \, = 1.393\, \pm\, 0.481$ for BL Lacs.
The probability for the two distributions
to be  from the same parent  distribution is
$p =  7.77\times 10^{-16}$.

\subsection{Correlations}

From the available data:
 photon flux $(F)$,
 photon Spectral Index ($\alpha_{\rm ph}$),
 and variability spectral index (${V\!I}$),
 we can make mutual correlations.

\begin{figure}
\begin{center}
\includegraphics[angle=0,width=10cm]{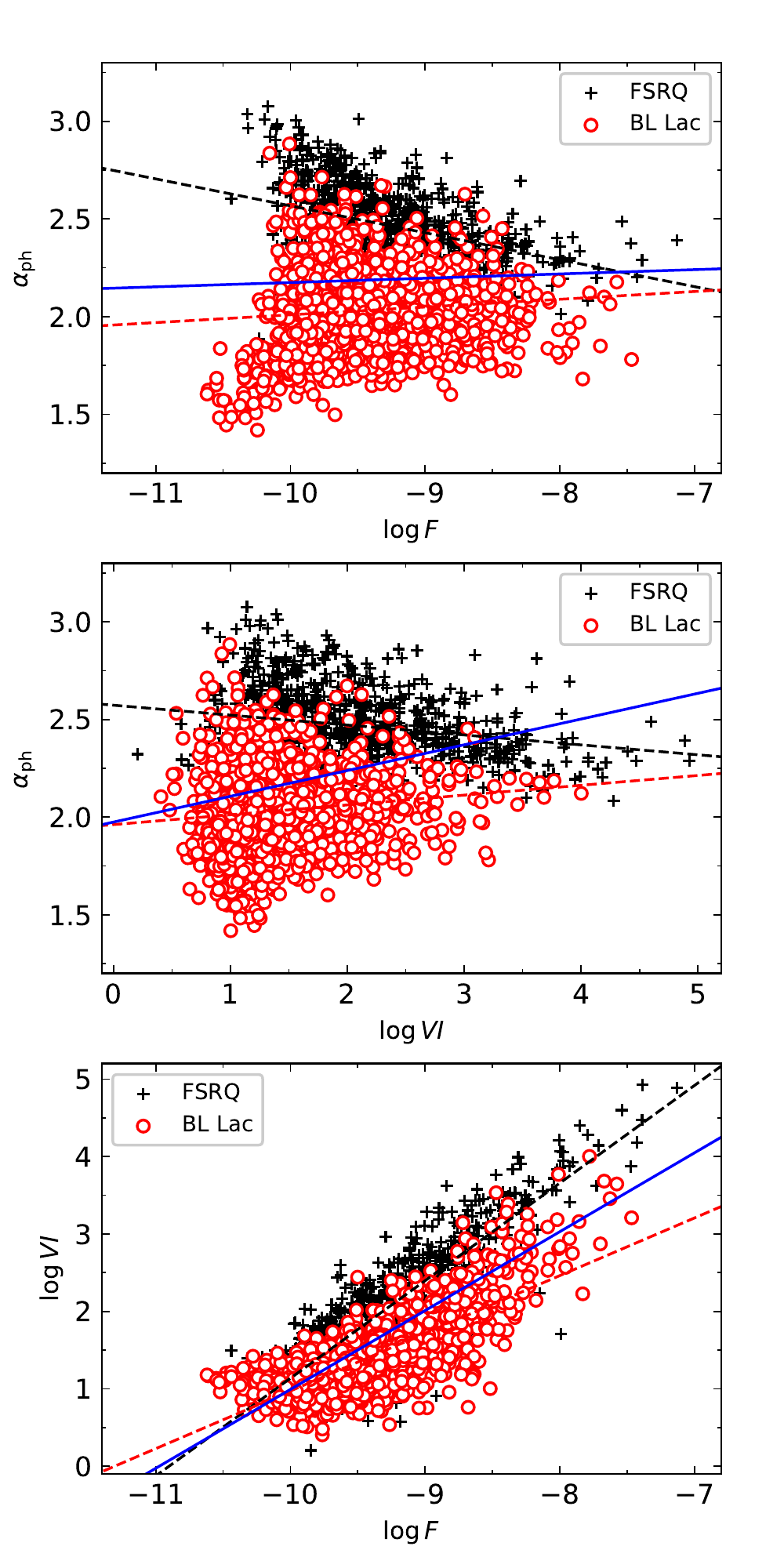}
\caption{Plot of mutual correlations.
  Symbol `plus' is for FSRQs, and
  `open circle' for BL Lacs.
  The straight blue line stands for the best fitting result for blazars (BL Lacs and FSRQs),
  `broken black line' for FSRQs, and
  `broken red line' for BL Lacs.
Upper panel is for the plot of photon spectral index ($\alpha_{\rm ph}$) versus  $\gamma$-ray photon flux (${\rm log}\,F$ (photon/cm$^2$/s));
middle panel for photon spectral index ($\alpha_{\rm ph}$) against variability index (${\rm log}\,{V\!I}$),
lower panel for variability index (${\rm log}\, {V\!I}$) against $\gamma$-ray photon flux (${\rm log}\, F$ (photon/cm$^2$/s)).
 } \label{Fig.Corr}
\end{center}
\end{figure}

 \textbf{\textit{Photon Flux versus Photon Spectral Index $(F - \alpha_{\rm ph})$}}:
    From the given $\gamma$-ray photon flux and the photon spectral index
   from the 4FGL catalogue,
   we investigated their mutual correlation and obtained
    $$\alpha_{\rm ph}\, = (0.022 \pm 0.012)\, {\rm log}\, F \, + 2.395 \pm 0.116$$
    with $r = 0.038$ and
    $p = 7.5\%$
    for  BL Lac and FSRQs.
    The corresponding best fitting result is shown in the upper panel of Fig. \ref{Fig.Corr}.
    When BL Lacs and FSRQs  are considered separately,
  one has
   $\alpha_{\rm ph}\, = -(0.138 \pm 0.012)\, {\rm log}\, F \, + 1.189 \pm 0.115$
    with $r = -0.369$ and $p = 4.42 \times10^{-27}$ for  FSRQs,
    and $\alpha_{\rm ph} \, = (0.040 \pm 0.012)\, {\rm log}\, F \, + 2.407 \pm 0.110$
    with $r = 0.09$ and  $p = 6.3\times 10^{-4}$  for BL Lacs.

   \textbf{\textit{Photon Spectral Index versus Variability Index ${\rm (} \alpha_{\rm {ph}} - {V\!I} {\rm )}$}}:
   The photon spectral index and variability index give following  linear  mutual correlation
   $$\alpha_{\rm ph}\, = (0.132 \pm 0.009)\, {\rm log}\, {V\!I} + 1.975 \pm 0.016$$
    with $r = 0.30$ and $p  = 8.3\times 10^{-48}$
    for BL Lacs and FSRQs,
    the corresponding best fitting result is shown in the middle panel of Fig. \ref{Fig.Corr}.
   We get following results  when BL Lacs and FSRQs are considered separately,
   $$\alpha_{\rm ph}\, = (0.050 \pm 0.012)\, {\rm log}\, {V\!I} + 1.962 \pm 0.017$$
    with $r = 0.114$ and $p  = 1.53 \times 10^{-5}$ for BL Lacs, and
   $$\alpha_{\rm ph}\, = - (0.051 \pm 0.009)\, {\rm log}\, {V\!I} + 2.573 \pm 0.020$$
    with $r = - 0.196$ and $p  = 2.65 \times 10^{-8}$ for FSRQs.

 \textbf{\textit{Flux versus Variability Index ${\rm (} F - {V\!I} {\rm )}$}}:
   From the  $\gamma$-ray photon flux  and the variability index,
   we obtained their mutual  correlation
    $${\rm log}\, {V\!I} \, = (1.018 \pm 0.018)\, {\rm log}\, F\, + 11.176 \pm 0.170$$
    with $r = 0.766$ and $p \sim 0$
    for  all BL Lacs and FSRQs.
    The corresponding best fitting result is shown in lower panel of Fig. \ref{Fig.Corr}.
  While for BL Lacs and FSRQs with available redshift,
   one has
   ${\rm log}\, {V\!I} \, = (0.745 \pm 0.018)\, {\rm log}\, F\, + 8.420 \pm 0.166$
   with $r = 0.746$ and $p \sim 0$ for  FSRQs, and
   ${\rm log}\, {V\!I} \, = (1.261 \pm 0.025)\, {\rm log}\, F\, + 13.741 \pm 0.230$
   with $r = 0.876$ and $p \sim 0$ for BL Lacs.

\subsection{Classifications}

From the mutual correlation analyses, it is found that BL Lacs and FSRQs show different correlation and they both occupy different regions in the plots.
In this sense, we can try to find a dividing line to separate BL Lacs and FSRQs, and further we can use this dividing line to tell BL Lacs from FSRQs when the BCUs are put in the plots.

In the last version of the Fermi catalogue \citep{Abd20,aje20}, there are 1517  blazar candidates unidentified type (BCUs).
It is interesting to divide them into BL Lacs and FSRQs.
In this work, we used a SVM, a kind of supervised machine learning (ML) method, to find a dividing line for separating the two blazar subclasses as in \citet{Yang2022b}.
SVM is widely used for classification and regression problems in astrophysics studies \citep{Xiao2022, Yang2022b, Wang2022, Solarz2017, Han2016}.
Consider there are two linearly separable samples in the $N$ dimensional parameter space, thus there are infinite numbers of $N-1$ dimensional hyperplanes can be found to separate them into two sides.
The SVM is, then, applied to determine the plane with the maximum margin, i.e., the maximum distance to the nearest samples.
For the case of non-linearly separable samples, SVM map the samples to a high-dimensional space and find the optimal separating hyperplane in the high-dimensional space.
The SVM requires a training data set and a data set, that randomly takes 70\% and 30\% sources of each type.
The training set is used to find the optimal hyperplane, the test set is used to evaluate the classification accuracy of the hyperplane.
In this work, we put BL Lac and FSRQ samples in the two-dimensional parameter space, that formed by either two (denote `$A$' and `$B$') of the three parameters ${\rm log}\,F$, $\alpha_{\rm ph}$ and ${\rm log}\,{V\!I}$.
Assuming the hyperplane, is a line in the two-dimensional space, is expressed as $w_{\rm 1}A+w_{\rm 2}B + m = 0$.
The factors $w_{\rm 1}$, $w_{\rm 2}$ and $m$ can be determined through training SVM with the training set. 
The $svm.LinearSVR$ (from $sklearn$ package) is employed as a SVM classifier, and the hyperparameters of $svm.LinearSVR$ need to be specified before the SVM training starts.
We iterate different combinations of hyperparameters in the training process until $w_{\rm 1}$, $w_{\rm 2}$ and $m$ converge to the the maximum margin.
At last, we get a number of different optimal dividing lines, and the one with the highest accuracy on the test set is the final optimal dividing line.

\begin{figure}[bht]
\includegraphics[width=1\columnwidth]{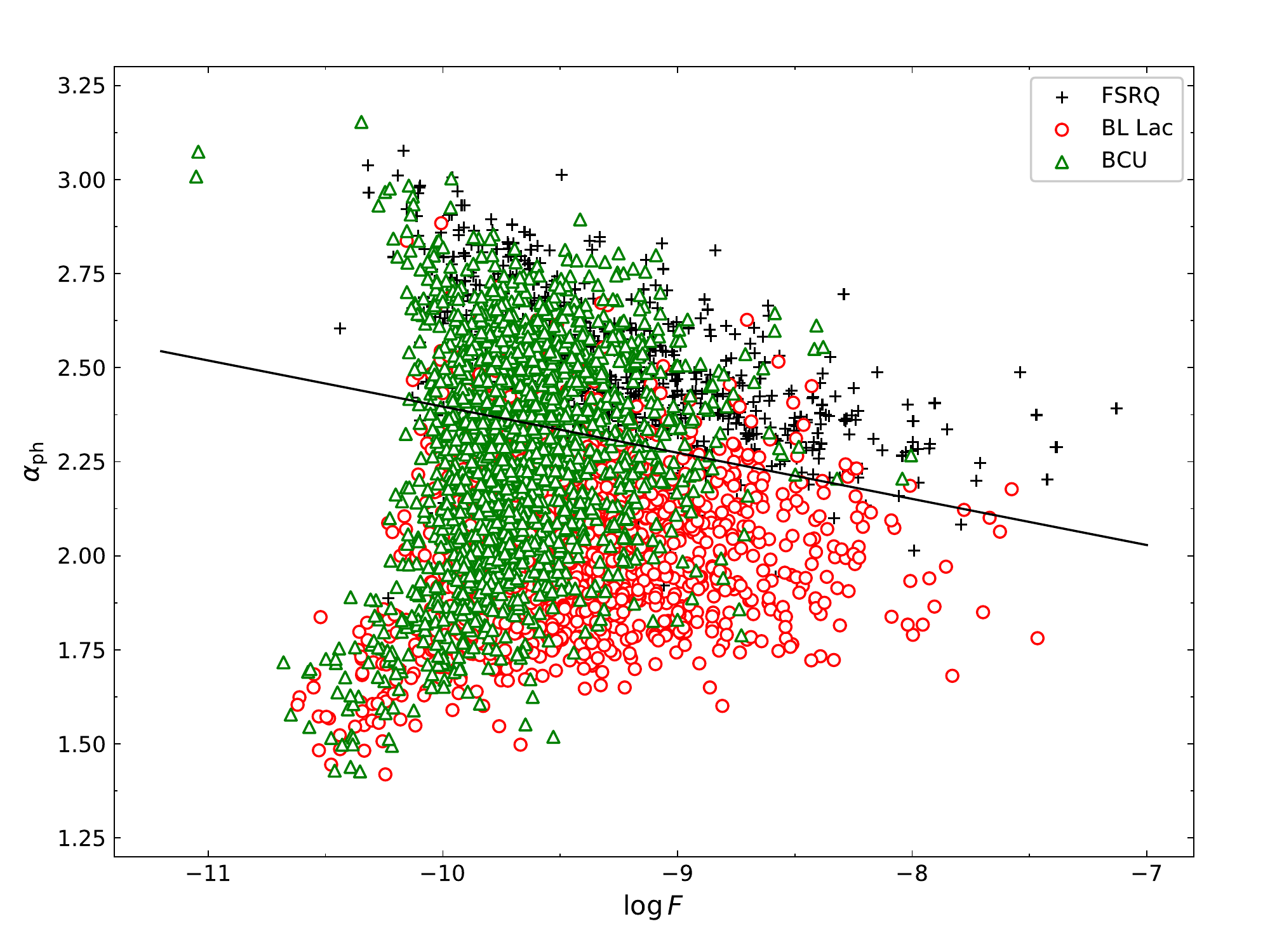}
\caption{Plot of photon spectral index ($\alpha_{\rm ph}$) against $\gamma$-ray photon flux (${\rm log}\,F$).
 Open circles stand for BL Lacs,
 plus for FSRQs, and
 triangle points for BCUs.
 The solid line ($\alpha_{\rm ph} = -0.223\,{\rm log}\,F + 1.170$) is obtained
 from the SVM method, it separates FSRQs and BL Lacs.}
\label{fig.a-F}
\end{figure}

When the SVM is adopted to the ${\rm (}F - \alpha_{\rm ph} {\rm )}$ data, the result gives an accuracy of 88.60\% for the separation and  a dividing line of $\alpha_{\rm ph} = -0.123\, {\rm log}\, F + 1.170$ as shown in Fig. \ref{fig.a-F}.
One can notice that FSRQs mainly occupy the region with $\alpha_{\rm ph} > -0.123\, {\rm log}\,F + 1.170$ and majority of BL Lacs occupy the region with $\alpha_{\rm ph} < -0.123\, {\rm log}\, F + 1.170$.

  When the 1517 BCUs are put into the plot,
  we found there are 639  BCUs locate in the region above the dividing line and they can be taken as FSRQ candidates (FC) while
  there are 878 BL Lac candidates (BC) since they are in the region below the dividing line.

\begin{figure}[bht]
\includegraphics[width=1\columnwidth]{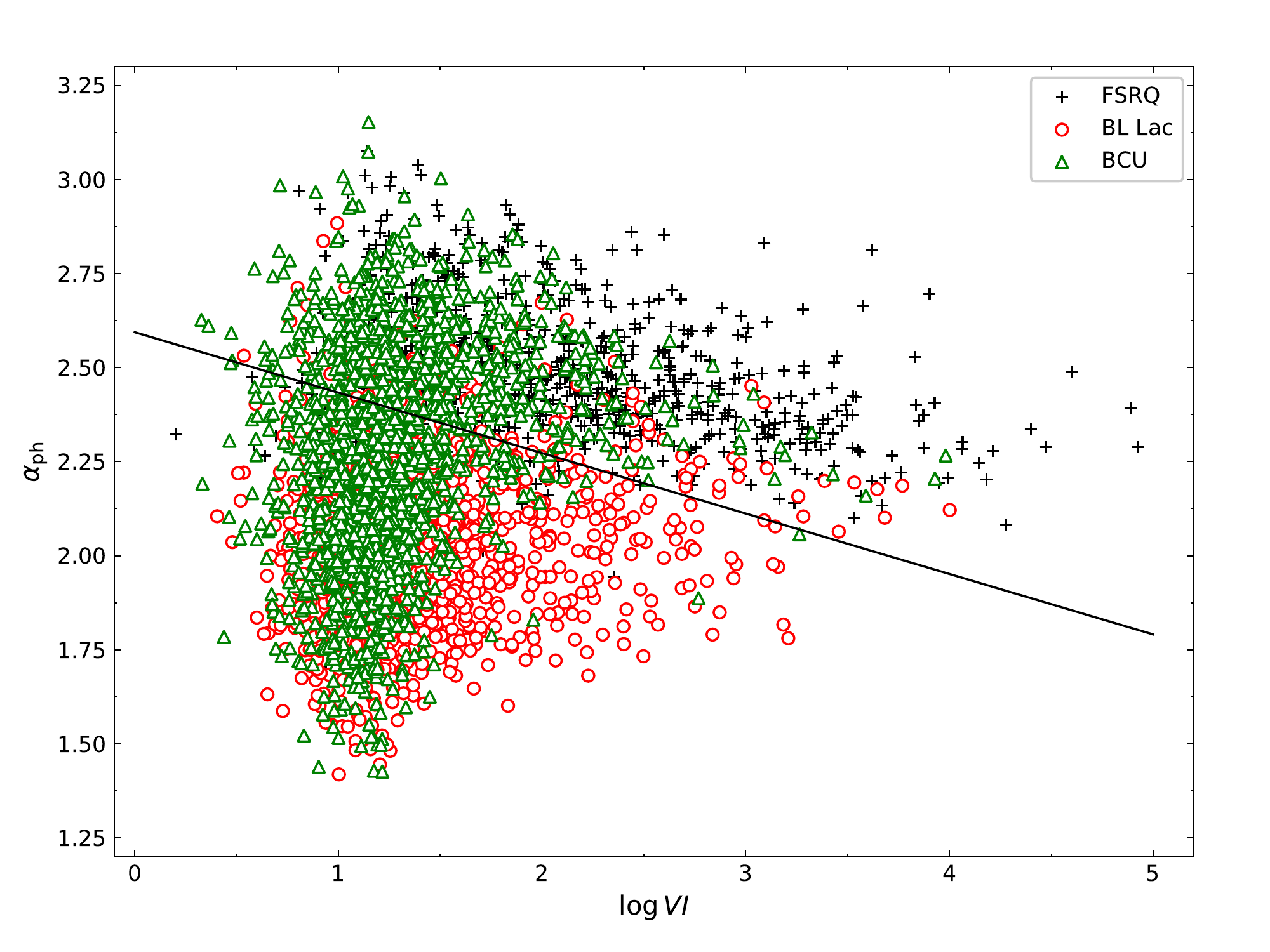}
\caption{Plot of variability index (${\rm log}\,{V\!I}$) against gamma-ray photon flux (${\rm log}\,F$).
 Open circles stand for BL Lacs,
 plus for FSRQs, and
 triangle points for BCUs.
 The solid line ($\alpha_{\rm ph} = -0.161\,{\rm log}\, {V\!I} + 2.594$) is obtained
 from the SVM method, it separates FSRQs and BL Lacs.}
\label{fig.a-VI}
\end{figure}

  When we considered the ${\rm (} \alpha_{\rm {ph}} - {V\!I} {\rm )}$ plot, we found that BL Lac and FSRQs can be divided by
      $\alpha_{\rm ph} = -0.161\, {\rm log}\, {V\!I} + 2.594$ with an  accuracy of 89.26\% as shown in Fig. \ref{fig.a-VI}.
    When the 1517 BCUs are put into the plot,
  we found
     585 FSRQ candidates (FC) and
    932 BL Lac candidates (BC).

\begin{figure}[bht]
\includegraphics[width=1\columnwidth]{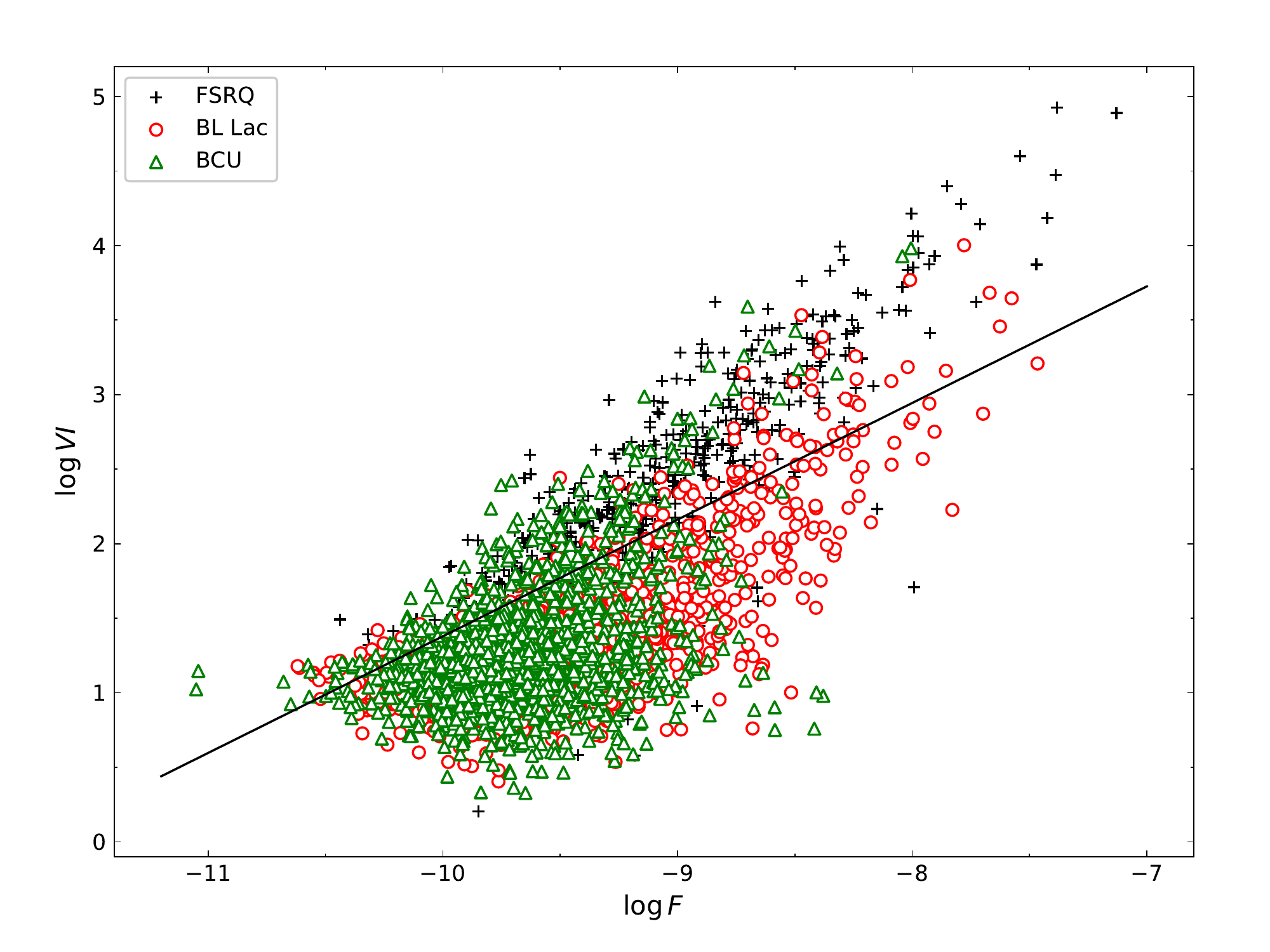}
\caption{Plot of variability index (${\rm log}\,{V\!I}$) against gamma-ray photon flux (log $F$).
 Open circles stand for BL Lacs,
 plus for FSRQs, and
 triangle points for BCUs.
 The solid line (${\rm log}\, {V\!I} = 0.792{\rm log}\, F + 9.203$) is obtained
 from the SVM method, it separates FSRQs and BL Lacs.}
\label{fig.VI-F}
\end{figure}

  For the  ${\rm (} F - {V\!I} {\rm )}$ plot, we found a dividing line of
  ${\rm log}\, {V\!I} = 0.792\,{\rm log}\,F + 9.203$ with an  accuracy of 79.16\% as in Fig. \ref{fig.VI-F}.
  Based on which we obtained
    337 FSRQ candidates (FC) and
    1180 BL Lac candidates (BC).

  In our consideration,
  we take a BCU as a BL Lac candidate (BC) if it is below the dividing line in the three plots,
  and as a possible BL Lac candidate (p-BC) if it is below the dividing line in any two plots;
  For FSRQs we also take the same consideration.
  Therefore,
  we have got 751 BL Lac candidates (BCs) and
  181 possible BL Lac candidates (p-BCs), namely 932 BC and p-BCs in total;
  210 FSRQ candidates (FCs)  and
  375 possible FSRQ candidates (p-FCs), namely 585 FCs and p-FCs in total.
  The ratio of the number of FCs and p-FCs versus BC and p-BC is
  $\sim {\frac{2}{3}}$ ( 585 versus 932 ).


  When we considered the ${\rm (} \alpha_{\rm {ph}} - {V\!I} {\rm )}$ plot, we found that BL Lac and FSRQs can be divided by
      $\alpha_{\rm ph} = -0.161\, {\rm log}\, {V\!I} + 2.594$ with an  accuracy of 89.26\% as shown in Fig. \ref{fig.a-VI}.
    When the 1518 BCUs are put into the plot,
  we found
     377 FCs and
    714 BCs.
  For the  ${\rm (} F - {V\!I} {\rm )}$ plot, we found a dividing line of
  ${\rm log}\, {V\!I} = 0.792\,{\rm log}\,F + 9.203$   with an  accuracy of 79.16\% as in Fig. \ref{fig.VI-F}.
  Based on which we obtained
  223 FCs and
    888 BCs.

  In our consideration,
  we take a BCU as a BL Lac candidate if it is below the dividing line in the three plots,
  and as a possible BL Lac candidate if it is below the dividing line in any two plots;
  For FSSRQs we also take the same consideration.
  Therefore,
  we have got 590 BCs and
  124 p-BCs,
  714 BC and p-BCs in total;
  145 FCs and
  232 p-FCs, 377 FC and p-FCs in total.


\section{Discussions}

   After the launch of \textit{Fermi}/LAT in 2008,
   a series of  catalogues have been released.
   The latest catalogue
   \citep{Abd20} contain 3743 blazars  and BCUs
    including 1432 BL Lacs, 794 FSRQs and 1517 BCUs.
    Recently,
    \citet{Yang2022b} studied the variability properties and the
    classification of BCUs based on a sample selected from the common blazars
    from \citep{fan16} and \citep{Abd20} with ${V\!I} > 18.48$.
    In the work  \citep{Abd20},
    who pointed out that  ${V\!I} > 18.48$ suggests a  source is variable.
    So, it is not necessary for one to take the restriction of ${V\!I} > 18.48$ since
    not all sources are violently variable.

    It is clear that there a lot of blazars with ${V\!I} < 18.48$,
    we will miss classification for a lot of BCUs if we only consider the
    BCUs with ${V\!I} > 18.48$.
    That is why we considered all BCUs listed in \citet{Abd20} in the present work.

\subsection{The average values}

    For the observational data,
    the $\gamma$-ray photon flux (${\rm log}\, F$),
    the photon spectral index ($\alpha_{\rm ph}$), and
    the variability index (${\rm log}\, {V\!I}$).
    It  can be found that the averaged values of three physics parameters
    in FSRQs are greater than those of BL Lacs.
    The K-S test indicates that the probability ($p$) for
    the distribution for FSRQs and that for BL Lacs to be from
    the same parent distribution is  $p < 6.708 \times 10^{-7}$.

    For the subclasses of BL Lacs,
    we will investigate the LBLs and HBLs,
    we do not consider IBLs since IBLs maybe include some LBLs and HBLs.
    In this sense, it is found that,
for the photon spectral index,
     $\langle \alpha_{\rm ph} \rangle = 2.197 \pm 0.168 $ for LBLs, and
     $\langle \alpha_{\rm ph} \rangle = 1.902 \pm 0.149 $ for HBLs,
     which show clear difference between LBLs and HBLs with
     $p = 5.5\times 10^{-92}$;
for the photon flux,
     $\langle {\rm log}\, F \rangle = -9.301 \pm 0.482 $ for LBLs, and
     $\langle {\rm log}\, F \rangle = -9.398 \pm 0.485 $ for HBLs
     with $p = 4.25\%$; and
for the variability index,
     $\langle {\rm log}\, {V\!I} \rangle = 1.584 \pm 0.594 $ for LBLs, and
     $\langle {\rm log}\, {V\!I}  \rangle = 1.377 \pm 0.421 $ for HBLs
     with $p = 1.33\times 10^{-6}$.

    When we considered FSRQs and LBLs for comparison,
    it is found that the probability for the two subclasses of blazars to
    be from the same parent population is
    $p = 1.75\times 10^{-75}$ for the photon spectral index, $\alpha_{\rm ph}$,
    $p = 37.8\%$ for the photon flux, ${\rm log}\, F$, and
    $p = 1.17\times 10^{-20}$ for variability index, ${\rm log}\, {V\!I}$.
    The comparisons between FSRQs and HBLs give
    $p \sim 0 $ for the photon spectral index, $\alpha_{\rm ph}$,
    $p = 4.6\times 10^{-3}$ for the photon flux, ${\rm log}\, F$, and
    $p = 9.39\times 10^{-56}$ for variability index, ${\rm log}\, {V\!I}$.

One can see  a clear difference in the photon spectral index ($\alpha_{\rm ph}$)
  between FSRQs and LBL, between FSRQ and HBL, and between LBLs and HBLs,
  giving
   $ \langle \alpha_{\rm ph}  \rangle |_{\rm FSRQ}  > \langle \alpha_{\rm ph}  \rangle |_{\rm LBL}  > \langle \alpha_{\rm ph}  \rangle |_{\rm HBL}$.
  It is also found that
  $\langle {\rm log}\,{V\!I}  \rangle |_{\rm FSRQ}  > \langle {\rm log}\, {V\!I}  \rangle |_{\rm LBL}  > \langle {\rm log}\, {V\!I} \rangle|_{\rm HBL}$.
  A  clear difference in photon flux (${\rm log}\, F$) between FSRQ and HBL ($p = 4.6\times 10^{-3}$) and
     a marginal different between HBLs and LBLs with a $p = 4.26\%$ are found.
  However,  no clear difference in the photon flux between LBL and FSRQs.
   We can say there is a sequence from FSRQ to LBL to HBL for photons spectral index and variability index that are similar to that pointed out by \citet{Fossati1998}, also see in \citet{Ghisellini2017}.

\subsection{The correlations for FSRQs and BL Lacs}

In this work, we considered the mutual correlations amongst $\alpha_{\rm ph} $, ${\rm log}\, F$, and ${\rm log}\, {V\!I}$ between FSRQs and BL Lacs.
There is a tendency for a positive correlation between spectral index ($\alpha_{\rm ph}$) and photon flux (${\rm log}\, F$) for known blazars.
When we considered BL Lacs and FSRQs separately, a clear anti-correlation is found for FSRQs with $p = 4.42\times 10^{-27}$ and a positive correlation for BL Lacs as shown in the upper panel of Fig. \ref{Fig.Corr}.
BL Lacs and FSRQs show different spectral index dependence on photon flux.
The both subclasses also show different spectral index dependence on the variability flux in the middle panel of Fig. \ref{Fig.Corr}, which indicates that the spectral index ($\alpha_{\rm ph}$) in FSRQs decreases with variability  index ${\rm log}\, {V\!I}$ while that in BL Lacs increases with variability index.
While for photon flux (${\rm log}\, F$) and variability index (${\rm log}\, {V\!I}$), blazars and the two subclasses all show positive correlation indicating that the variability index (${\rm log}\, {V\!I}$) increases with photon flux (${\rm log}\, F$) as shown in the lower panel of Fig. \ref{Fig.Corr}.
    
We need to note that the linear correlations that we obtained amongst $\alpha_{\rm ph} $, ${\rm log}\, F$, and ${\rm log}\, {V\!I}$ do not mean strict mathematic linear correlation but demonstrate possible trends between two parameters.
It is more reasonable to explore trends instead strict mathematical linear correlations.
The theoretical relationship between the three parameters (and also for other astronomical quantities) is rarely investigated.
Because these observational quantities show a significant discrepancy, the discrepancy leaves enough space for various models to explain the phenomenon.
In the case of individual sources, the observational discrepancy can come for several reasons, the source's intrinsic reason (e.g. the black hole mass and spin, the accretion ratio, etc) and the external reason (e.g. the gas and dust density of the host galaxy, the magnetic field, the distance, etc).
In the case of many sources, the distribution of the sources may serve a selection effect of the telescope or a few sources in the universe that have very high/low values of some quantities.
All the above-mentioned reasons could obstruct us from discovering the linear correlation mathematically.
For the sources in our sample, it is natural that most of the sources have low photon flux (i.e. ${\rm log}\,F < -8.5$) and the variability index (i.e. ${\rm log}\, {V\!I} < 2.5$) and those sources with very high photon flux (i.e. ${\rm log}\,F > -7.5$) and the variability index (i.e. ${\rm log}\, {V\!I} > 4.5$) are very rare in the universe, see in the middle and lower panels of Fig \ref{Fig.Corr}.

Through the correlations study amongst $\alpha_{\rm ph} $, ${\rm log}\, F$, and ${\rm log}\, {V\!I}$ in this work.
We conclude that the FSRQs show trends of anti-correlation for $\alpha_{\rm ph} \, {vs} \, {\rm log}\, F$ and $\alpha_{\rm ph} \, {vs} \, {\rm log}\, {V\!I}$, while the BL Lacs show trends of positive correlation for the two, see in the upper and middle panels of Fig \ref{Fig.Corr}.
And both FSRQs and BL Lacs show trend of positive correlation for ${\rm log}\, {V\!I} \, {vs} \, {\rm log}\, F$.

\subsection{The classification for BCUs}

   It is found that most of  BL Lacs and FSRQs occupy different regions in the
   plots of
   $\alpha_{\rm ph} $ versus ${\rm log}\, F$ (Fig. \ref{fig.a-F}),
   $\alpha_{\rm ph} $ versus ${\rm log}\, {V\!I}$ (Fig. \ref{fig.a-VI}), and
   ${\rm log}\, {V\!I}$ versus  ${\rm log}\, F$ (Fig. \ref{fig.VI-F}).
   When the support vector machine (SVM) method is adopted to the relevant data,
   separating lines were obtained, which can be used to
   give BL Lac and FSRQ candidates when BCUs are put in to the
   plots.

   In those cases,
    we have obtained 639 FCs and 878 BCs
    in the $\alpha_{\rm ph} $ versus ${\rm log}\, F$ plot (Fig. \ref{fig.a-F}),
    582 FCs and 932 BCs in the
   $\alpha_{\rm ph} $ versus ${\rm log}\, {V\!I}$ plot (Fig. \ref{fig.a-VI}), and
    337 FCs and 1180 BCs in the
   ${\rm log}\, {V\!I}$ versus  ${\rm log}\, F$ plot (Fig. \ref{fig.VI-F}).

    We take  a BCU  as an FC
     if it is classified as an FC in all the three cases, and
     we take a BCU  as a p-FC if it is classified as an FC in any two cases.
     We also give the similar considerations for BCs and p-BCs.
     Our candidate classifications are shown in Table  \ref{Com}.
     Therefore, we have 932 BC and p-BCs,
      585 FCs and p-FCs  giving a
       number ratio $\sim {\frac{2}{3}}$.

     We also made comparison with the classification results in
      \citet{kang19}.
      We  take a BCU as an FC if it was classified as an FC in all their three
      considerations,
      and as a p-FC if it was classified as an FC in any two of their considerations  \citep{kang19}.
      For BC, we also give a similar classification.
      If this case, we obtained that there are
        302 FCs and 109 p-FCs,
        556 BCs and 114 p-BCs, which are given in Col. (6) in Table  \ref{Com}.
  When we  compared our classifications with those in \citet{kang19},
     it is  found that there are  1091 common sources and 426 sources in the present work were not included in \citet{kang19}.
     For the  1091 common sources, our analyses give
       590 BCs in this work (TW) correspond to 492 BCs, 10 FCs, 68 p-BCs, and 20 p-FCs in the work \citet{kang19},
       145 FCs in TW correspond to 4 BCs, 119 FCs, 6 p-BCs, and 16 p-FCs in Kang,
       124 p-BCs in TW correspond to 60 BCs, 15 FCs, 24 p-BCs, and 25 p-FCs in Kang, and
       231 p-FCs in TW correspond to 10 BCs, 147 FCs, 16 p-BCs, and 48 p-FCs in Kang.
     Therefore,
      our 590 BCs and 124 p-BCs (total 714) correspond to 552 BCs and 92 p-BCs (total 644) in Kang,
      it means that out of 714 sources in our considerations,
       644  are the similar to that by Kang giving a 90.2\% goodness of fit,
      and our 145 FCs and 231 p-FCs (total 376) correspond to 266 FCs and 64 p-FCs  (total 330) in Kang giving a 87.8\% goodness of fit.



\begin{table*}
\scriptsize \caption{Classification for the BCU sources in this work}
\label{Com} \centering
\begin{tabular}{lcccccccc}
\hline \hline
4FGL name &
${\rm log}\, F$ &
${\rm log}\, {V\!I}$ &
$\alpha_{\rm ph}$ &
Class$^{\alpha_{\rm ph}-F}$ &
Class$^{\alpha_{\rm ph}-{V\!I}}$ &
Class$^{F-{V\!I}}$ &
Class-TW  &
Class(K19) \\
(1)&(2)&\,(3)&(4)&\,\,(5)&(6)&(7)&(8)&(9)\\
\hline
4FGL J0001.2+4741	&	-9.900 	&	1.403 	&	2.272 	&	BL Lac	&	BL Lac	&	FSRQ	&	P-B		&	BL Lac	\\
4FGL J0001.6-4156	&	-9.549 	&	1.421 	&	1.775 	&	BL Lac	&	BL Lac	&	BL Lac	&	BL Lac	&	BL Lac	\\
4FGL J0001.8-2153	&	-10.043 	&	1.390 	&	1.877 	&	BL Lac	&	BL Lac	&	FSRQ	&	P-B		&	NN		\\
4FGL J0002.1-6728	&	-9.587 	&	1.098 	&	1.848 	&	BL Lac	&	BL Lac	&	BL Lac	&	BL Lac	&	BL Lac	\\
4FGL J0002.3-0815	&	-9.924 	&	1.114 	&	2.092 	&	BL Lac	&	BL Lac	&	BL Lac	&	BL Lac	&	NN		\\
4FGL J0002.4-5156	&	-10.108 	&	1.248 	&	1.914 	&	BL Lac	&	BL Lac	&	FSRQ	&	P-B		&	NN		\\
4FGL J0003.1-5248	&	-9.463 	&	0.903 	&	1.916 	&	BL Lac	&	BL Lac	&	BL Lac	&	BL Lac	&	BL Lac	\\
4FGL J0003.3-1928	&	-9.372 	&	1.698 	&	2.282 	&	BL Lac	&	BL Lac	&	BL Lac	&	BL Lac	&	P-F		\\
4FGL J0003.3-5905	&	-9.916 	&	1.006 	&	2.274 	&	BL Lac	&	BL Lac	&	BL Lac	&	BL Lac	&	P-B		\\
4FGL J0003.5+0717	&	-9.814 	&	1.039 	&	2.217 	&	BL Lac	&	BL Lac	&	BL Lac	&	BL Lac	&	NN		\\
4FGL J0007.7+4008	&	-9.351 	&	1.552 	&	2.140 	&	BL Lac	&	BL Lac	&	BL Lac	&	BL Lac	&	BL Lac	\\
4FGL J0008.0-3937	&	-9.920 	&	1.220 	&	2.626 	&	FSRQ	&	FSRQ	&	BL Lac	&	P-F		&	FSRQ	\\
4FGL J0008.4+1455	&	-9.286 	&	1.715 	&	2.079 	&	BL Lac	&	BL Lac	&	BL Lac	&	BL Lac	&	BL Lac	\\
...             & ...       & ...  & ...       & ...       & ...       & ...       & ...    & ...   \\
...             & ...       & ...  & ...       & ...       & ...       & ...       & ...    & ...   \\
...             & ...       & ...  & ...       & ...       & ...       & ...       & ...    & ...   \\
...             & ...       & ...  & ...       & ...       & ...       & ...       & ...    & ...   \\
\hline \hline
\end{tabular}
\\This table is available in its entirety in
machine-readable forms.
\end{table*}

\section{Conclusions}

 In this work,
   3743 blazars from the 4FGL catalogue \citep{Abd20, aje20},
   which includes 1432 BL Lacs, 794 FSRQs, and 1517 BCUs.
   We analyzed their averaged values and mutual correlation amongst the photon spectral index, variability index and the photon flux for
    the known blazars. SVM method is adopted to separate BL Lacs and FSRQs, afterwards, we used the separating line to classify the the BCUs into BL Lac and FSRQs. We also proposed to classify a BCU as a BL Lac object if it is classified in all the three cases,
          and as a possible BL Lac candidate if it is classified as a BL Lac in only two cases.
          For FSRQ candidate, we also take the similar considerations.
          Our classifications are compared with those by Kang et al. (2019).
    Our conclusions are as follows:

     1. The $\gamma$-ray photon flux, spectral index, and variability index of FSRQs are
        higher than those of BL Lacs  for the known blazar sample.
        There is a sequence from FSRQs to LBLs to HBLs being similar to that in \citet{Fossati1998}.

     2.  A positive correlation is found
           between $\gamma$-ray flux and the photon spectral index for the whole sample,
           but an anti-correlation is found for FSRQs and a positive correlation for BL Lacs.
            In addition, a positive correlation is found
           between variability index (${\rm log}\, {V\!I}$) and the $\gamma$-ray photon spectrum index($\alpha_{\rm ph}$)
           for the whole sample, but an anti-correlation for FSRQs and a positive correlation for BL Lacs.
           We think those two positive correlations for the whole sample are  apparent.

     3.  We adopted the SVM machine learning method to classify  BL Lacs and FSRQs in
          the $\alpha_{\rm ph}~ {\rm v.s.}~ {\rm log}\, F$,
         and $\alpha_{\rm ph}~ {\rm v.s.}~ {V\!I}$ plots and
         ${\rm log}\, F ~ {\rm v.s.}~ {V\!I}$.
         We obtained 932 BL Lac candidates and possible BL Lac candidates, and
          585 FSRQ candidates and possible FSRQ candidates.

     4. We compared our classifications with those in  \citet{kang19} and found that
         for the common sources, there is a goodness fit of 90.2\% for BL Lac and possible BL Lac candidates, and
         a goodness of 87.8\% for FSRQ and possible FSRQ candidates with  those by
         \citet{kang19}.

\vspace{6pt} 

\authorcontributions{Conceptualization, J.H.F. and K.Y.C.; methodology, J.H.F. and H.B.X.; writing---original draft preparation. J.H.F., W.X.Y., J.C.L., and G.H.C.; visualization and discussion, J.H.Y., Y.H.Y., and D.X.W. 
All authors have read and agreed to the published version of the manuscript.}

  \funding{The work is partially supported by the National Natural Science
  Foundation of  China (NSFC U2031201, NSFC 11733001, U2031112, NSFC 12133004, NSFC 12103012),
  Guangdong Major Project of Basic and Applied Basic Research (Grant No. 2019B030302001).
  We also acknowledge the science research grants from the China
  Manned Space Project with NO. CMS-CSST-2021-A06, and the
  supports for Astrophysics  Key Subjects of Guangdong Province and Guangzhou City.
  The work is also supported by Guangzhou University (YM2020001).}
  \institutionalreview{Not applicable.}
  \dataavailability{} 
  \acknowledgments{}
  \conflictsofinterest{The authors declare no conflict of interest.}

\begin{adjustwidth}{-\extralength}{0cm}
\reftitle{References}

\end{adjustwidth}

\begingroup
\scriptsize 
\begin{longtable}{lcccccccc}
\caption{Classification for the BCU sources in this work} \\ \hline \hline
\centering
4FGL name &
${\rm log}\, F$ &
${\rm log}\, VI$ &
$\alpha_{\rm ph}$ &
Class$^{\alpha_{\rm ph}-F}$ &
Class$^{\alpha_{\rm ph}-VI}$ &
Class$^{F-VI}$ &
Class-TW  &
Class(K19) \\
(1)&(2)&\,(3)&(4)&\,\,(5)&(6)&(7)&(8)&(9) \\ \hline
4FGL J0001.2+4741	&	-9.900 	&	1.403 	&	2.272 	&	BL Lac	&	BL Lac	&	FSRQ	&	P-B	&	BL Lac	\\	
4FGL J0001.6-4156	&	-9.549 	&	1.421 	&	1.775 	&	BL Lac	&	BL Lac	&	BL Lac	&	BL Lac	&	BL Lac	\\	
4FGL J0001.8-2153	&	-10.043 	&	1.390 	&	1.877 	&	BL Lac	&	BL Lac	&	FSRQ	&	P-B	&	NN	\\	
4FGL J0002.1-6728	&	-9.587 	&	1.098 	&	1.848 	&	BL Lac	&	BL Lac	&	BL Lac	&	BL Lac	&	BL Lac	\\	
4FGL J0002.3-0815	&	-9.924 	&	1.114 	&	2.092 	&	BL Lac	&	BL Lac	&	BL Lac	&	BL Lac	&	NN	\\
\end{longtable}
\endgroup

\label{lastpage}


\begin{thebibliography}{99}

\bibitem[Abdollahi et al.\@(2020)]{Abd20}  Abdollahi, S., Acero, F., Ackermann, M., et al. 2020, ApJS, 247, 35

\bibitem[Acero et al.\@(2015)]{acero15} Acero, F., Ackermann, M., Ajello, M., et al., 2015, ApJS, 218, 23

\bibitem[Ajello et al. (2020)]{aje20} Ajello, M., Angioni, R., Axelsson, M., et al. 2020, ApJ, 892, 105


\bibitem[Fan et al.(2021)]{fan21} Fan, J. H., et al. 2021, ApJ, 253, 10

\bibitem[Ghisellini et al.(2014)]{ghi14} Ghisellini, G., Tavecchio, F., Maraschi, L., Celotti, A., \& Sbarrato, T. 2014, Nature, 515, 376


\bibitem[Stickel et al.(1991)]{sti91} Stickel, M., Padovani, P., Urry, C. M., Fried, J. W., \& K\"uhr, H. 1991, ApJ, 374, 431

\bibitem[Urry  \& Padovani(1995)]{urry95} Urry, C. M., \& Padovani, P. 1995, PASP, 107, 803

\bibitem[Wills et al. \@(1992)]{Wills92} Wills, B. et al. 1992, ApJ,398,454





\bibitem[Yang et al.(2022b)]{Yang2022b} Yang, W. X., Xiao, H.B., Wang, H. G., Liu, Y. et al. 2022, RAA, in press




\bibitem[Zhou et al. \@(2021)]{zhou21} Zhou, R. X., Zheng, Y. G., Zhu, K. R., Kang, S. J. 2021, ApJ, 915, 59

\bibitem[Villata et al. (2006)]{Villata2006} Villata, M., Raiteri, C. M., Balonek, T. J., et al. 2006, A\&A, 453, 817

\bibitem[Gupta et al. (2016)]{Gupta2016} Gupta, A. C., Agarwal, A., Bhagwan, J., et al. 2016, MNRAS, 458, 1127

\bibitem[Lister et al. (2018)]{Lister2018} Lister, M. L., Aller, M. F., Aller, H. D., et al. 2018, ApJS, 397 234, 12,

\bibitem[Lister et al. (2021)]{Lister2021} Lister M. L., Homan D. C., Kellermann K. I., Kovalev Y. Y., Pushkarev A. B., Ros E., Savolainen T., 2021, ApJ, 923, 30

\bibitem[Lind \& Blandford (1985)]{Lind85} Lind, K.R. \& Blandford, R.D. 1985, ApJ, 295, 358

\bibitem[Stocke et al. (1990)]{Stocke1990} Stocke, J. T., Morris, S. L., Gioia, I., et al. 1990, ApJ, 348, 141

\bibitem[Scarpa et al. (1997)]{Scarpa1997} Scarpa, R., \& Falomo, R. 1997, A\&A, 325, 109



\bibitem[Fan \& Xie (1996)]{Fan96} Fan, J. H. \& Xie, G.Z., 1996, A\&A, 306, 55


\bibitem[Padovani \& Giommi(1995)]{pg95}  Padovani, P., \& Giommi, P. 1995, ApJ, 444, 567

\bibitem[Nieppola et al. (2006)]{nie06} Nieppola, E., Tornikoski, M., \& Valtaoja, E. 2006, A\&A, 445, 441

\bibitem[Abdo et al.(2010a)]{abdo10} Abdo, A. A., et al. 2010a, ApJ, 716, 30

\bibitem[Fan et al.(2016)]{fan16} Fan, J. H., et al. 2016, ApJ, 226, 20


\bibitem[Nolan et al. (2012)]{Nolan12} Nolan, P. L., Abdo, A. A., Ackermann, M., et al. 2012, ApJS, 199, 31

\bibitem[Ballet et al. \@(2020)]{Ballet2020} Ballet, J., Burnett, T. H., Digel, S. W., \& Lott, B. 2020, arXiv e-prints, arXiv:2005.11208


\bibitem[Hassan et al.\@(2013)]{Hassan2013} Hassan, T., Mirabal, N., Contreras, J. L., \& Oya, I. 2013, MNRAS, 428, 220

\bibitem[Doert et al.\@(2014)]{Doert2014} Doert, M., \& Errando, M. 2014, ApJ, 782, 41

\bibitem[Chiaro et al.\@(2016)]{Chiaro2016} Chiaro, G., Salvetti, D., La Mura, G., et al. 2016, MNRAS, 462, 3180

\bibitem[Parkinson et al.\@(2016)]{Parkinson2016} Saz Parkinson, P. M., Xu, H., Yu, P. L. H., et al. 2016, ApJ, 820, 8

\bibitem[Lefaucheur \& Pita (2017)]{Lefaucheur2017} Lefaucheur, J., \& Pita, S. 2017, A\&A, 602, A86

\bibitem[Yi et al.\@(2017)]{Yi2017} Yi, T. F., Zhang, J., Lu, R. J., Huang, R., \& Liang, E. W. 2017, ApJ, 838, 34

\bibitem[Bai et al.\@(2018)]{Bai2018} Bai, Y., Liu, J. F., \& Wang, S. 2018, RAA, 18, 118

\bibitem[Ma et al.\@(2019)]{Ma2019} Ma, Z., Xu, H., Zhu, J., et al. 2019, ApJS, 240, 34

\bibitem[Kang et al.(2019a)]{Kang2019} Kang, S. J., Fan, J. H., Mao, W., et al. 2019, ApJ, 872, 189


\bibitem[Kang et al.(2019b)]{kang19} Kang, S. J., Li, E. Z., Ou,W. J.,et al., 2019, ApJ, 887, 134

\bibitem[Fossati et al.(1998)]{Fossati1998} Fossai, G., Maraschi, L., Celotti, A., Comastri, A., Ghisellini, G., 1998, MNRAS, 299, 433

\bibitem[Ghisellini et al. (207)]{Ghisellini2017} Ghisellini, G., Righi, C., Costamante, L., Tavecchio, F. 2017, MNRAS, 469, 255

\bibitem[Xiao et al. (2022)]{Xiao2022} Xiao, H.B., Zhu, J. T., Fu, L. P., Zhang, S. H., Fan, J. H. 2022, PASJ, 0, 1

\bibitem[Wang et al. (2022)]{Wang2022} Wang, C., Bai, Y., et al. 2022, A\&A, 659, 144

\bibitem[Solarz et al. (2017)]{Solarz2017} Solarz, A., Bilicki, M., Gromadzki, M., et al. 2017, A\&A, 606, 39 

\bibitem[Han et al. (2016)]{Han2016} Han, B., Ding, H. P ., et al. 2016, RAA, 16, 74

\end{thebibliography}
\end{document}